\documentclass[conference]{IEEEtran}
\IEEEoverridecommandlockouts
\usepackage{cite}
\usepackage{amsmath,amssymb,amsfonts}
\usepackage{graphicx}
\usepackage{textcomp}
\usepackage{xcolor}
\usepackage{booktabs}
\usepackage{array}
\usepackage{url}
\usepackage{algorithm}
\usepackage{algpseudocode}
\usepackage{hyperref}
\hypersetup{colorlinks=true,linkcolor=blue,citecolor=blue,urlcolor=blue}

\begin{document}

\title{When the Ruler is Broken: Parsing-Induced Suppression 
in LLM-Based Security Log Evaluation}

\author{
\IEEEauthorblockN{Chaitanya Vilas Garware}
\IEEEauthorblockA{
\textit{Department of Computer and Information Sciences}\\
\textit{University of Alabama at Birmingham}\\
Birmingham, AL, USA\\
chaitanyagarware7@gmail.com
}
\and 
\IEEEauthorblockN{Sharif Noor Zisad}
\IEEEauthorblockA{
\textit{Department of Computer and Information Sciences}\\
\textit{University of Alabama at Birmingham}\\
Birmingham, AL, USA\\
mrzisad@gmail.com
}
}

\maketitle

\begin{abstract}

LLM-based SOC log classifiers are commonly evaluated using regular-expression
pipelines that extract structured fields from free-form model output.
We demonstrate that this practice introduces a class of silent, systematic
evaluation errors, which we term \textit{parsing-induced suppression} that
can cause a fully functional model to appear completely non-functional.
Using OpenSOC-AI, a LoRA fine-tuned TinyLlama-1.1B system for
security log threat classification, as a reproducible case study, we show that
a strict regex parser reported \textbf{0\% threat accuracy} while a corrected
fuzzy parser recovered \textbf{76\% threat accuracy} on the same model outputs
and the same evaluation set. A gap of 76 percentage points attributable
entirely to evaluation methodology.
Severity accuracy remained constant at 58\% under both parsers,
providing a built-in control that isolates field name format mismatch
as the causal mechanism rather than model degradation.
For external reference, Claude Sonnet evaluated zero-shot on the same
50 example set achieved 88\% threat accuracy and 58\% severity accuracy
under the same fuzzy protocol.
Residual errors under fuzzy evaluation concentrate in three categories including reconnaissance, brute force, and credential stuffing, each contributing
all 4 misclassifications, a pattern that reflects class-boundary
difficulty among behaviorally adjacent log types rather than global model failure.
We propose \textit{SOC-Bench v0}, a benchmark framework comprising a
standardized 13 category threat taxonomy, minimum statistical power
requirements, fuzzy field extraction specification, and a public scoring
script intended to prevent parser specific accuracy distortion in future
SOC LLM research.A central finding of this study is that, in SOC LLM evaluation, the parser plays a critical role in the evaluation process. Silent parser failures can cause the reported accuracy to reflect characteristics of the evaluation pipeline rather than the actual performance of the model.

\end{abstract}


\section{Introduction}

Security operation centers face a fundamental throughput problem.
Modern infrastructure generates log volumes that exceed the capacity
of human analysts to triage manually, and this gap is widest in
resource-constrained organizations that lack dedicated security personnel.
Large language models have attracted substantial attention as a potential
solution, with recent work demonstrating that even billion-scale models
fine-tuned on modest labeled datasets can produce structured threat
classifications from raw log entries~\cite{opensoc,ferrag}.
Lightweight fine-tuning on domain-specific data
can adapt a general-purpose model to produce MITRE-mapped threat labels,
severity scores, and remediation recommendations with minimal compute. However, an LLM-based SOC classifier is only as useful as the evaluation
that tells us whether it works.
Unlike computer vision benchmarks, where standardized datasets and
deterministic scoring produce comparable results across research groups,
LLM evaluation in the security domain relies on a pipeline step that
receives almost no methodological scrutiny: the parser that converts
free-form model output into structured labels before scoring.

Models fine-tuned on structured output templates do not always
reproduce those templates.
Instruction-tuned base models carry formatting preferences from their
pretraining that can produce field names in title case (\texttt{Threat Type:})
where the fine-tuning template expected all-caps snake\_case
(\texttt{THREAT\_TYPE:}).
A regex extractor that searches for \texttt{THREAT\_TYPE} will silently
discard every prediction where the model used \texttt{Threat Type},
scoring the example as incorrect regardless of whether the semantic
content of the prediction was right.
The model is not failing. The ruler is broken. The distinction matters because deployment decisions, benchmarking conclusions, and research directions may all be distorted by evaluation artifacts rather than underlying model capability.
This paper makes that failure mode concrete, reproducible, and quantifiable.

We conduct a methodological audit of OpenSOC-AI~\cite{opensoc},
a LoRA fine-tuned TinyLlama-1.1B system for SOC log threat classification.
The original system reported 68\% threat accuracy under its evaluation pipeline.
When we apply a strict regex replication of that pipeline to a retrained
checkpoint, we observe 0\% threat accuracy. Every prediction discarded
before scoring while a fuzzy parser applied to the same outputs recovers
76\% accuracy.
Severity accuracy, extracted by a pattern that happens to match the model's
output format, remains stable at 58\% under both parsers.

Our contributions are as follows:

\begin{itemize}
\item \textbf{Parsing-induced suppression, defined and quantified.}
We introduce the term \textit{parsing-induced suppression} to describe
the class of evaluation errors arising from field-name format mismatch
between fine-tuning templates and model output.
We demonstrate a 76 percentage-point accuracy gap attributable to this
failure on a 50-example held-out evaluation set, with a built-in control
(unchanged severity accuracy) that isolates parsing as the causal variable.

\item \textbf{Failure concentration analysis.}
Residual errors under corrected evaluation are not uniformly distributed.
All 12 remaining misclassifications occur in three behaviorally adjacent
categories (Reconnaissance, Brute Force, Credential Stuffing),
10 of 13 categories achieved 100\% accuracy under fuzzy evaluation; however, several categories contained only a small number of evaluation samples, limiting statistical confidence in per-class estimates.
This concentration pattern distinguishes systematic class-boundary
difficulty from general model failure and motivates targeted
rather than global improvement.

\item \textbf{External baseline comparison.}
We evaluate Claude Sonnet zero-shot on the same evaluation set
under the same fuzzy protocol, providing a large-model reference point
that contextualizes the fine-tuned small model's performance.

\item \textbf{SOC-Bench v0.}
We propose a benchmark framework comprising a 13-category MITRE-aligned
threat taxonomy, minimum per-class evaluation size requirements,
fuzzy field extraction specification, and a public scoring script
to enable comparable, parser-agnostic evaluation across research groups.
\end{itemize}

The remainder of this paper is organized as follows.
Section~\ref{sec:related} reviews related work.
Section~\ref{sec:system} describes the OpenSOC-AI reference system.
Section~\ref{sec:pipeline} explains the strict and fuzzy parser designs.
Section~\ref{sec:setup} details the experimental setup.
Section~\ref{sec:results} presents results.
Section~\ref{sec:failure} analyzes residual failure concentration.
Section~\ref{sec:socbench} introduces SOC-Bench v0.
Section~\ref{sec:discussion} discusses broader implications.
Section~\ref{sec:limitations} states limitations.
Section~\ref{sec:conclusion} concludes.

\section{Background and Related Work}
\label{sec:related}

\subsection{LLMs for Cybersecurity and SOC Automation}

The application of large language models to cybersecurity tasks has
expanded significantly in recent years.
Ferrag et al.~\cite{ferrag} survey LLM applications across vulnerability
detection, malware analysis, intrusion detection, and log analysis,
documenting accuracy gains on structured classification tasks.
Prior cybersecurity NLP work has shown that domain-specific language modeling can improve security text understanding tasks, but these studies generally focus on model performance rather than auditing the extraction pipelines used to compute reported metrics.
More recently, OpenSOC-AI~\cite{opensoc} applied parameter-efficient
fine-tuning to the SOC log triage problem, showing that a 1.1B parameter
model fine-tuned on 450 labeled examples can produce structured threat
classifications competitive with larger baselines.

These works share a common evaluation pattern: accuracy is measured by
extracting structured fields from free-form model output using
regular expressions or template matching, then comparing against
ground truth labels.
The extraction step is rarely described in methodological detail,
and its failure modes are not reported.
The present work fills that gap.

\subsection{Parameter-Efficient Fine-Tuning}

Low-Rank Adaptation (LoRA)~\cite{lora} and its quantized variant
QLoRA~\cite{qlora} have substantially reduced the compute requirements
for domain adaptation of large language models.
By introducing trainable low-rank matrices into attention and MLP layers
while keeping base model weights frozen, LoRA enables fine-tuning on
consumer hardware with minimal labeled data.
OpenSOC-AI~\cite{opensoc} applied QLoRA to TinyLlama-1.1B~\cite{tinyllama},
achieving structured output from a model that produces no structured
output in its base form.
We use the same fine-tuning configuration in this work.

\subsection{Evaluation Methodology and Reproducibility}

Evaluation fragility is a documented concern in adjacent fields.
In adversarial machine learning, inconsistent threat models and
evaluation protocols produce results that are not comparable across
papers~\cite{ferrag}.
In NLP, the reproducibility crisis has motivated shared evaluation
infrastructure and standardized scoring~\cite{augsurvey}.
In computer vision, benchmark datasets and deterministic metrics
enable meaningful cross-system comparison.

SOC LLM evaluation has none of these.
Published accuracy figures are produced by systems that vary in
parser implementation, field naming conventions, normalization logic,
and evaluation set size.
A model that achieves 76\% under a fuzzy parser may appear to achieve 0\%
under a strict parser applied to the same outputs as this paper demonstrates.
Without evaluation methodology transparency, published figures cannot
be meaningfully compared.

\subsection{OpenSOC-AI as the Base System}

This work is an evaluation-focused extension of OpenSOC-AI~\cite{opensoc}.
We do not modify the model architecture, training data, or fine-tuning
procedure described in that work.
Our contribution is the methodological audit: we apply two different
evaluation pipelines to the same model outputs, characterize the
discrepancy, attribute its cause, analyze the residual failure pattern,
and propose evaluation infrastructure to prevent recurrence.

\section{Reference System: OpenSOC-AI}
\label{sec:system}

OpenSOC-AI~\cite{opensoc} fine-tunes TinyLlama-1.1B-Chat-v1.0~\cite{tinyllama}
using QLoRA~\cite{qlora} on security log classification.
The configuration targets all seven attention and MLP projection layers
at rank $r = 16$ with scaling $\alpha = 32$, dropout 0.05, trained for
three epochs on a single NVIDIA T4 GPU.
Training completes in approximately five minutes on 450 labeled examples.

The model takes a security log entry as input and produces structured
free-form text containing: threat type, MITRE technique ID,
MITRE tactic, severity, risk score, evidence summary,
and remediation recommendations.
The dataset distribution across the evaluation set is shown in
Table~\ref{tab:dist}.

\begin{table}[ht]
\caption{Evaluation Set Distribution (from OpenSOC-AI~\cite{opensoc})}
\label{tab:dist}
\centering
\begin{tabular}{lc}
\toprule
\textbf{Threat Category (Broad)} & \textbf{Eval Examples} \\
\midrule
SQL Injection (all variants) & 8 \\
Data Exfiltration & 7 \\
Windows Threat (all variants) & 5 \\
DDoS / Denial of Service & 4 \\
Path / Directory Traversal & 4 \\
Reconnaissance / Scanning & 4 \\
Brute Force Attack & 4 \\
Credential Stuffing & 4 \\
SSH Brute Force & 4 \\
Local File Inclusion & 3 \\
No Threat / Normal Traffic & 1 \\
Command Injection & 1 \\
Cross-Site Scripting & 1 \\
\midrule
\textbf{Total} & \textbf{50} \\
\bottomrule
\end{tabular}
\end{table}

The original evaluation pipeline reported 68\% threat accuracy,
precision 0.71, recall 0.66, F1 0.68, and severity accuracy 58\%.
MITRE technique ID evaluation was excluded from the original report
because the extractor failed to reliably match technique identifiers
across model output format variation~\cite{opensoc}, an early
indicator of the parsing fragility this paper investigates.

\section{Evaluation Pipeline and Parser Design}
\label{sec:pipeline}

\subsection{The Problem: Free-Form Output Meets Rigid Extraction}

Fine-tuned instruction models do not always reproduce the exact
formatting of their training templates.
Base models arrive with formatting preferences established during
pretraining and instruction tuning: TinyLlama-1.1B consistently
generates field names in title case with spaces
(\texttt{Threat Type:}, \texttt{Severity:}, \texttt{MITRE Technique ID:})
while the OpenSOC-AI fine-tuning template uses all-caps
snake\_case (\texttt{THREAT\_TYPE:}, \texttt{SEVERITY:}, \texttt{MITRE\_ID:}).

A regex extractor searching for \texttt{THREAT\_TYPE} with case-folding
still fails to match \texttt{Threat Type} because the space-versus-underscore
distinction is not addressed by case-insensitive matching alone.
The extractor silently drops the prediction.
No exception is raised. No flag is set.
The example is scored as a misclassification.

\subsection{Strict Parser}

The strict parser follows the original OpenSOC-AI evaluation pipeline.
It applies regular expressions keyed to exact field name strings:

\begin{algorithm}
\caption{Strict Field Extraction}
\begin{algorithmic}[1]
\For{key \textbf{in} [\texttt{THREAT\_TYPE}, \texttt{SEVERITY}, \texttt{MITRE\_ID}]}
  \State match $\leftarrow$ \texttt{re.search(rf"\{key\}[:\textbackslash s]+([`\textbackslash n,]+)", text, IGNORECASE)}
  \If{match}
    \State fields[key] $\leftarrow$ match.group(1).strip().upper()
  \EndIf
\EndFor
\end{algorithmic}
\end{algorithm}

This parser correctly extracts \texttt{SEVERITY} because the model
generates \texttt{Severity:} and the \texttt{IGNORECASE} flag handles
the case difference.
It fails to extract \texttt{THREAT\_TYPE} because the model generates
\texttt{Threat Type:} and the underscore-to-space difference is not
covered by any flag.
The asymmetric extraction behavior, success on severity, failure on
threat type, is the mechanism that produces 0\% threat accuracy
alongside 58\% severity accuracy.

\subsection{Fuzzy Parser}

The fuzzy parser replaces exact-match patterns with format-tolerant
regular expressions that handle the full range of field name variants
observed in TinyLlama-1.1B output:

\begin{algorithm}
\caption{Fuzzy Field Extraction}
\begin{algorithmic}[1]
\Require Raw model output string $T$
\Ensure Structured fields: \textsc{ThreatType}, \textsc{Severity}, \textsc{Mitre}
\State \textbf{Step 1 — Loop truncation:}
\State \quad $T \leftarrow T[\ :\ \texttt{index}(\texttt{``\#\#\# Input:''})\ ]$ if marker found
\State
\State \textbf{Step 2 — Fuzzy key matching (per field):}
\State \quad \textsc{ThreatType}: match \textit{Threat[space/\_/-]*Type} + separator
\State \quad \textsc{Severity}: match \textit{Severity} + separator
\State \quad \textsc{Mitre}: match \textit{MITRE[space/\_]*Technique[*][ID]} + separator
\State \quad Separators accepted: \texttt{:} \texttt{-} \texttt{=} with optional whitespace
\State
\State \textbf{Step 3 — Value normalization:}
\State \quad Map \textsc{ThreatType} $\rightarrow$ SOC-Bench canonical category
\State \quad \quad e.g., \textit{``sql injection -- union''} $\rightarrow$ \texttt{SQL INJECTION}
\State \quad Map \textsc{Severity} $\rightarrow$ \{CRITICAL, HIGH, MEDIUM, LOW\}
\State
\State \Return \{\textsc{ThreatType}, \textsc{Severity}, \textsc{Mitre}\}
\end{algorithmic}
\end{algorithm}

The fuzzy parser accepts all observed field name variants including:
\texttt{THREAT\_TYPE:}, \texttt{Threat Type:}, \texttt{Threat-Type:},
\texttt{threat\_type:}, and \texttt{Threat\_Type:}.
It applies broad-category normalization so that fine-grained model
outputs (e.g., \texttt{SQL Injection -- OS Command via SQLi}) are
mapped to canonical evaluation categories (e.g., \texttt{SQL INJECTION})
before scoring.

\textbf{Critical point.}
The fuzzy parser does not change the model.
It does not modify model weights, training data, inference parameters,
or output logits.
It changes only how the model's existing outputs are interpreted.
Any accuracy difference between the two parsers is attributable
entirely to the evaluation pipeline.

\section{Experimental Setup}
\label{sec:setup}

\subsection{Model Training}

We retrained the OpenSOC-AI model from scratch following the configuration
in~\cite{opensoc}: QLoRA applied to TinyLlama-1.1B-Chat-v1.0,
rank $r=16$, $\alpha=32$, seven target modules, three training epochs,
NVIDIA T4 GPU.
Training completed in 4.5 minutes with a final training loss of 0.089,
consistent with the original report.
LoRA adapters were saved separately from the base model weights;
the base model was not modified.

The training set contains 450 labeled security log examples covering
12 broad threat categories.
The evaluation set contains 50 held-out examples covering 13 broad
categories under SOC-Bench normalization, with per-category counts
shown in Table~\ref{tab:dist}.

\subsection{Inference Configuration}

Inference used \texttt{max\_new\_tokens=120}, \texttt{do\_sample=False},
and \texttt{pad\_token\_id=eos\_token\_id}. Input prompts were truncated
to \texttt{max\_length=400} tokens to prevent overflow.
Outputs were post-processed with loop truncation before field extraction.

\subsection{Evaluation Conditions}

Three evaluation conditions were applied:

\textbf{Condition 1: Strict parser.}
The original OpenSOC-AI pipeline applied to the retrained checkpoint.
Exact field name matching with IGNORECASE, no normalization.

\textbf{Condition 2: Fuzzy parser.}
Format-tolerant extraction with broad-category normalization
applied to the same model outputs as Condition 1.
Model weights are identical; only the parser differs.

\textbf{Condition 3: Claude Sonnet zero-shot.}
Claude Sonnet (Anthropic, 2025) was evaluated zero-shot on the same
50-example evaluation set using a structured-output system prompt
specifying the same 13 threat categories used in SOC-Bench normalization.
Model outputs were scored using the same fuzzy normalization logic as
Condition 2, ensuring parser configuration does not confound the
model comparison.
Claude Sonnet is a large-parameter proprietary model;
it was not fine-tuned on security logs for this evaluation.
It serves as a large-model reference point, not a fine-tuned baseline.

\section{Results}
\label{sec:results}

\subsection{Parser Comparison}

Table~\ref{tab:parsers} shows the primary result: strict versus fuzzy
parser applied to the same TinyLlama-1.1B + LoRA outputs
on the 50-example evaluation set.

\begin{table}[ht]
\caption{Strict vs.\ Fuzzy Parser Comparison ($N = 50$, Same Model Outputs)}
\label{tab:parsers}
\centering
\begin{tabular}{lccc}
\toprule
\textbf{Metric} & \textbf{Strict} & \textbf{Fuzzy} & \textbf{Difference} \\
\midrule
Threat Accuracy   & 0.0\%  & 76.0\% & \textbf{+76.0pp} \\
Severity Accuracy & 58.0\% & 58.0\% & 0.0pp \\
\bottomrule
\end{tabular}
\end{table}

The result is stark. Under the strict parser, threat accuracy is 0\%:
not a single threat type prediction is extracted, because every instance
of \texttt{Threat Type:} is silently discarded.
Under the fuzzy parser, 76\% of predictions are correctly extracted
and matched.
The 76 percentage-point gap is attributable entirely to parsing-induced
suppression.

The severity row is the control.
Severity accuracy is 58\% under both parsers, because the model's
\texttt{Severity:} output format happens to match the strict extractor's
\texttt{SEVERITY} pattern under case folding.
No change in model behavior occurred between the two conditions.
The only change was the evaluation pipeline.
This control isolates field-name format mismatch as the causal mechanism.

\begin{figure}[ht]
\centering
\includegraphics[width=\columnwidth]{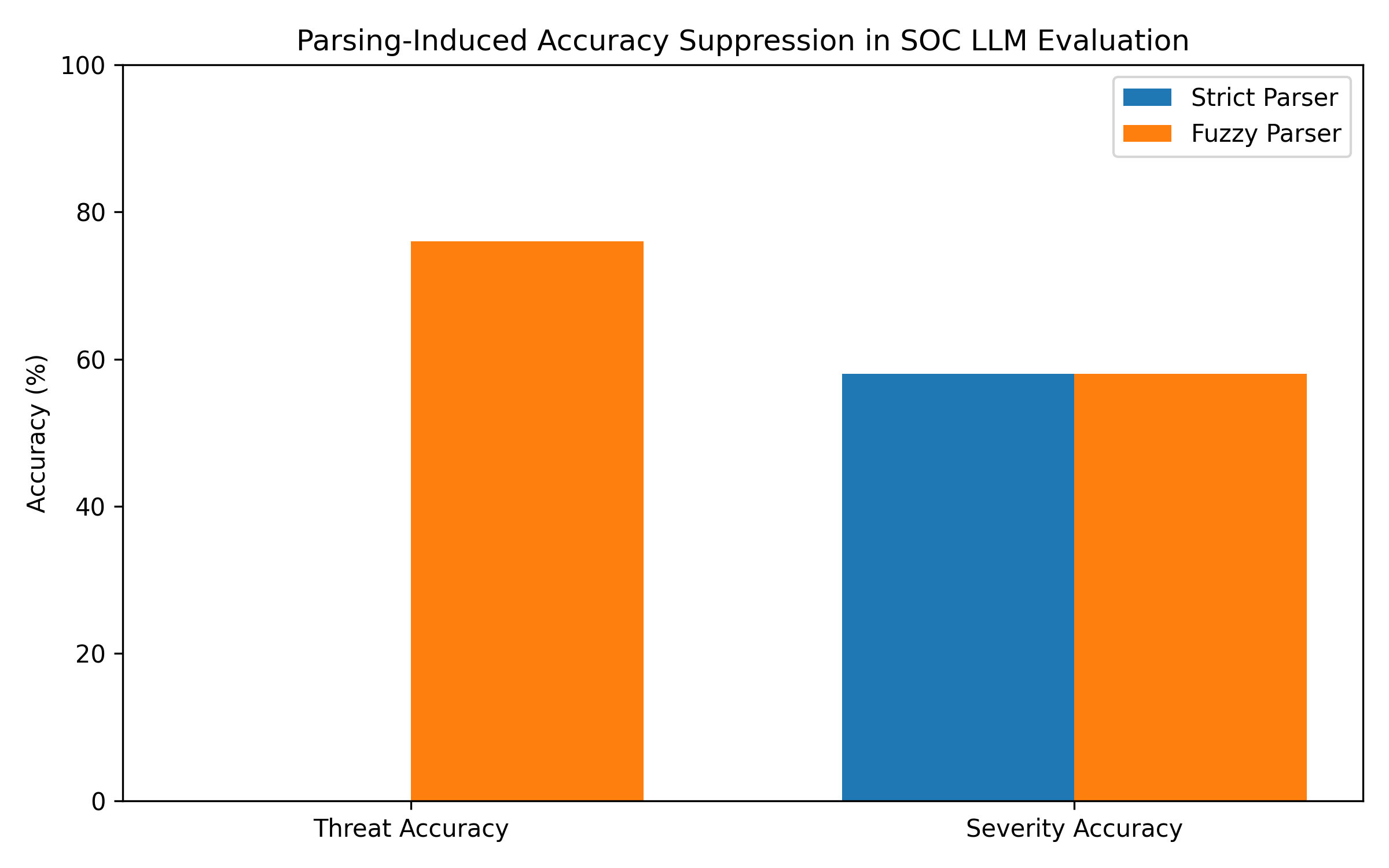}
\caption{Strict vs.\ fuzzy parser comparison on TinyLlama-1.1B + LoRA ($N=50$).
Threat accuracy rises from 0\% (strict) to 76\% (fuzzy) on identical model outputs.
Severity accuracy remains unchanged at 58\% under both parsers,
serving as a built-in control that isolates field-name format mismatch
as the sole causal variable. The 76pp gap is parsing-induced suppression,
not model failure.}
\label{fig:parser}
\end{figure}

Figure~\ref{fig:parser} illustrates this result.
The visual makes the control interpretation immediately clear:
a parser that recovers threat accuracy from 0\% to 76\% while
leaving severity accuracy unchanged is demonstrating a parser-specific
failure, not a model-specific one.

\subsection{Model Comparison}

\begin{figure}[ht]
\centering
\includegraphics[width=\columnwidth]{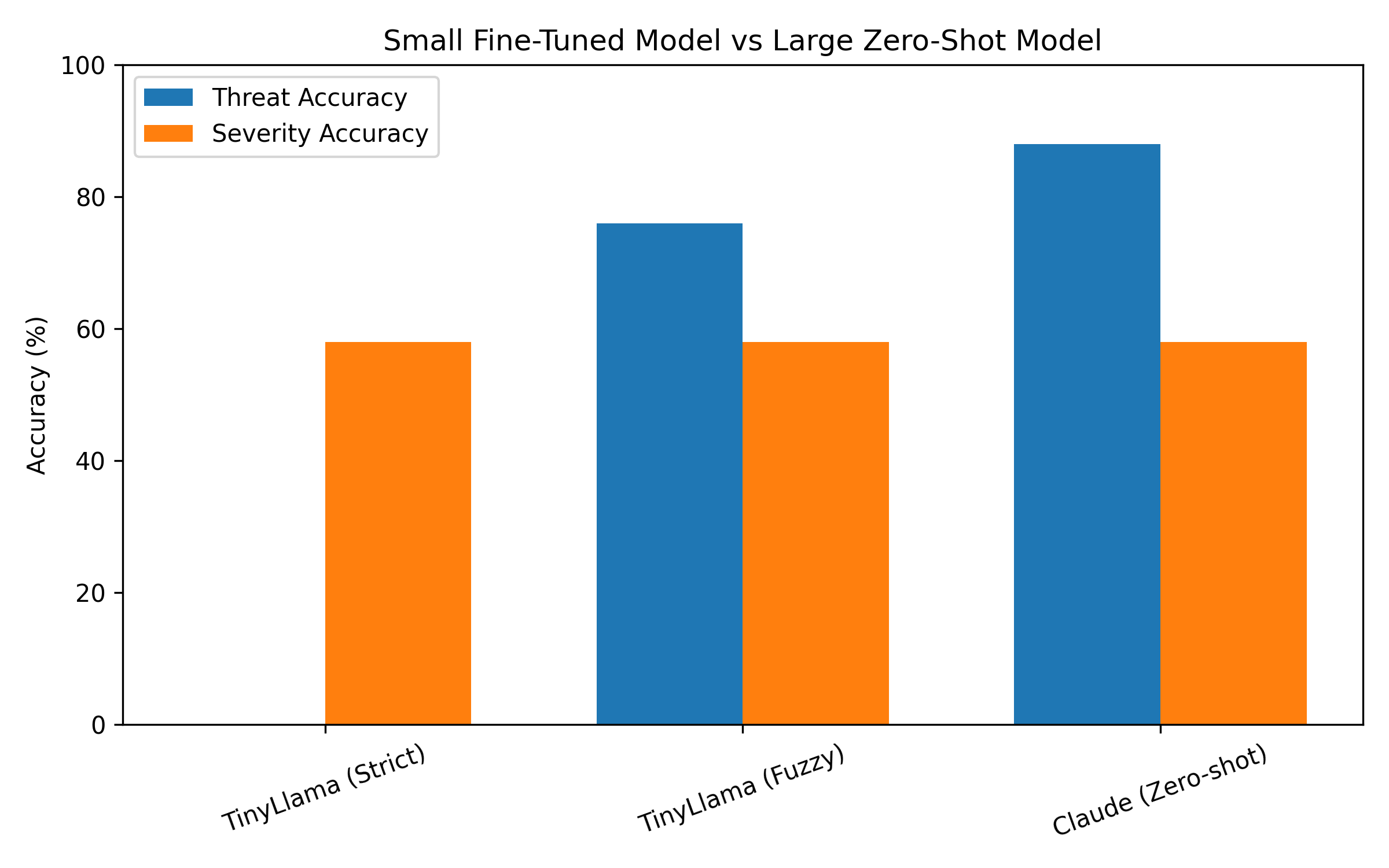}
\caption{Cross-model comparison under identical fuzzy evaluation protocol ($N=50$).
TinyLlama-1.1B + LoRA (fuzzy) reaches 76\% threat accuracy vs.\ Claude Sonnet
zero-shot at 88\%, a 12pp gap, not the 88pp gap reported by the strict parser.
All three conditions produce identical severity accuracy (58\%),
confirming the severity metric is parser-agnostic and model-driven.}
\label{fig:models}
\end{figure}

Table~\ref{tab:models} and Figure~\ref{fig:models} place the fine-tuned
model in context alongside the large-model zero-shot baseline.

\begin{table}[ht]
\caption{Model Comparison ($N = 50$, Fuzzy Evaluation Protocol)}
\label{tab:models}
\centering
\begin{tabular}{lcc}
\toprule
\textbf{Model} & \textbf{Threat Acc.} & \textbf{Severity Acc.} \\
\midrule
TinyLlama-1.1B + LoRA (strict) & 0.0\%  & 58.0\% \\
TinyLlama-1.1B + LoRA (fuzzy)  & 76.0\% & 58.0\% \\
Claude Sonnet (zero-shot)        & 88.0\% & 58.0\% \\
\bottomrule
\end{tabular}
\end{table}

Several observations follow from Table~\ref{tab:models}.

First, the strict-parser result (0\%) would have led a researcher
to conclude that fine-tuning failed entirely and that TinyLlama is
unsuitable for SOC log classification.
The fuzzy-parser result (76\%) leads to the opposite conclusion:
fine-tuning produces a model that performs well on 10 of 13 threat
categories and demonstrates competitive performance relative to a substantially larger zero-shot baseline.
The evaluation pipeline is the difference between these interpretations.

Second, Claude Sonnet at 88\% threat accuracy establishes that the
SOC log classification task is learnable by capable models.
The 12pp gap between TinyLlama fuzzy (76\%) and Claude zero-shot (88\%)
is real and meaningful; we do not dismiss it.
It reflects residual model limitations concentrated in three specific
categories (Section~\ref{sec:failure}).
But the gap is 12 percentage points, not 88 percentage points.
The strict parser was reporting the latter.

Third, all three conditions produce identical severity accuracy (58\%).
This is not a coincidence: severity extraction was not affected by the
field-name format mismatch, so it is stable across all three conditions.
This stability is what makes severity a useful control in this experiment.

\section{Failure Concentration Analysis}
\label{sec:failure}

Under fuzzy evaluation, 12 of 50 examples remain misclassified (24\%).
These errors are not distributed across the threat taxonomy.
They are concentrated in three categories:
Reconnaissance, Brute Force, and Credential Stuffing,
each contributing 4 misclassifications.
All other categories achieve 100\% accuracy under the fuzzy parser.

\begin{figure}[ht]
\centering
\includegraphics[width=\columnwidth]{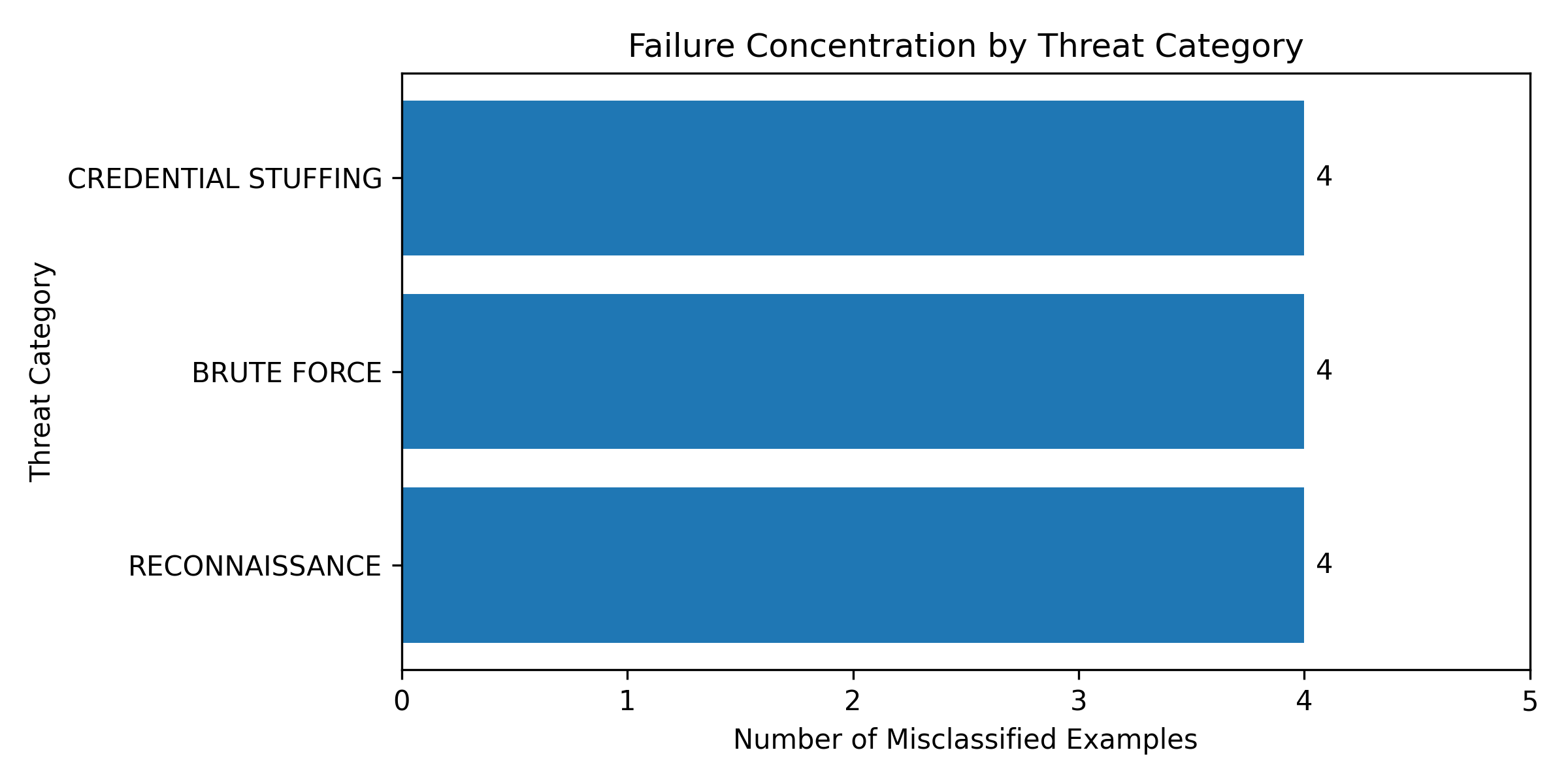}
\caption{Failure concentration under fuzzy evaluation.
All 12 residual errors occur exclusively in three behaviorally adjacent
categories (Reconnaissance, Brute Force, Credential Stuffing), each
contributing 4 misclassifications.
The remaining 10 categories achieve 100\% accuracy.
This concentration reflects systematic class-boundary difficulty
among high-frequency-pattern categories, not global model failure.}
\label{fig:failures}
\end{figure}

Table~\ref{tab:perclass} and Figure~\ref{fig:failures} show the per-class
results.

\begin{table}[ht]
\caption{Per-Class Accuracy Under Fuzzy Evaluation ($N = 50$)}
\label{tab:perclass}
\centering
\begin{tabular}{lcc}
\toprule
\textbf{Category} & \textbf{$n$} & \textbf{Accuracy} \\
\midrule
SQL Injection        & 8 & 100\% \\
Data Exfiltration    & 7 & 100\% \\
Windows Threat       & 5 & 100\% \\
DDoS                 & 4 & 100\% \\
Path Traversal       & 4 & 100\% \\
SSH Brute Force      & 4 & 100\% \\
LFI                  & 3 & 100\% \\
Command Injection    & 1 & 100\% \\
XSS                  & 1 & 100\% \\
No Threat            & 1 & 100\% \\
\midrule
Reconnaissance       & 4 & \textbf{0\%} \\
Brute Force          & 4 & \textbf{0\%} \\
Credential Stuffing  & 4 & \textbf{0\%} \\
\midrule
\textbf{Overall}     & \textbf{50} & \textbf{76\%} \\
\bottomrule
\end{tabular}
\end{table}

\subsection{Why These Three Categories?}

The concentration of errors in Reconnaissance, Brute Force, and
Credential Stuffing is not arbitrary.
These categories share substantial overlap in their observable log
characteristics:

\begin{itemize}
\item \textbf{Repeated connection patterns.}
Reconnaissance, brute force, and credential stuffing attacks
all manifest as high-frequency repeated requests from similar
source IP ranges, making source-based features non-discriminating.

\item \textbf{Authentication failure codes.}
Both brute force and credential stuffing produce repeated
401 HTTP status codes or SSH authentication failure messages.
Reconnaissance probes may produce 404 and 403 codes that
superficially resemble authentication-related responses.

\item \textbf{Automated user agents.}
All three categories are associated with automated tooling,
and their user agent strings such as curl, Python-requests, custom
scanning agents overlap significantly.

\item \textbf{High-frequency request timing.}
The timing patterns of scanning and authentication attacks
are similar at the log entry level, where individual entries
rather than sequences are classified.
\end{itemize}

These overlaps mean that even human SOC analysts frequently require
additional context beyond a single log entry to distinguish these
categories reliably.
The model's difficulty at these boundaries reflects a genuine
ambiguity in the classification task, not a general failure of
the fine-tuning approach.

\subsection{What the Concentration Pattern Does Not Mean}

We are careful not to overclaim from this result.
The per-class evaluation set sizes of 4 examples each are
below the 20-example minimum recommended by SOC-Bench v0.
The 95\% Wilson confidence intervals for 0/4 performance
span [0\%, 60\%], meaning the true per-class accuracy could
be anywhere in that range.
We observe 0\% across all three categories, which is consistent
with systematic failure, but we cannot establish the precise
failure rate from 4 examples.

We also observed generation artifacts including prompt-repetition loops
in a subset of raw model outputs for these categories.
However, we did not conduct controlled experiments to establish
whether these artifacts causally explain the failures, and we
do not claim that fixing generation parameters would produce 100\%
accuracy.
The remaining errors require further investigation on an expanded
evaluation set, with controlled inference parameter ablation,
before causal attribution is possible.

What we can say with confidence is that the failure pattern is
not uniform.
A model that achieves 100\% accuracy on SQL Injection, Data
Exfiltration, DDoS, Path Traversal, SSH Brute Force, and five
other categories is not globally failing.
Its errors are concentrated at a semantically coherent boundary,
which is precisely the information needed to direct improvement effort.

\section{SOC-Bench: A Reproducible Evaluation Framework}
\label{sec:socbench}

The failures documented in this paper are structural risks in any SOC LLM
evaluation that uses regex extraction from free-form output.
They are not quirks of OpenSOC-AI or TinyLlama.
Every research group building such a system faces the same pipeline
fragility, and currently there is no shared evaluation standard that
prevents it.
SOC-Bench v0 is proposed to fill that gap.

\subsection{Threat Taxonomy}

SOC-Bench defines 13 canonical threat categories aligned to
MITRE ATT\&CK v14~\cite{attack}, shown in Table~\ref{tab:taxonomy}.
Categories are defined at the broad level to enable fair comparison
across models trained on different label granularities.
Subcategory labels in training data are permitted;
scoring uses broad-category normalization before comparison.

\begin{table}[ht]
\caption{SOC-Bench v0 Canonical Threat Taxonomy}
\label{tab:taxonomy}
\centering
\begin{tabular}{ll}
\toprule
\textbf{ID} & \textbf{Category} \\
\midrule
SB-01 & SQL Injection \\
SB-02 & Cross-Site Scripting (XSS) \\
SB-03 & Command Injection \\
SB-04 & Path / Directory Traversal \\
SB-05 & Local File Inclusion (LFI) \\
SB-06 & Brute Force \\
SB-07 & Credential Stuffing \\
SB-08 & Reconnaissance / Scanning \\
SB-09 & Denial of Service / DDoS \\
SB-10 & Data Exfiltration \\
SB-11 & Lateral Movement / Privilege Escalation \\
SB-12 & Malware / C2 Activity \\
SB-13 & No Threat / Normal Traffic \\
\bottomrule
\end{tabular}
\end{table}

\subsection{Protocol Requirements}

A SOC-Bench-compliant evaluation must satisfy four requirements.

\textbf{R1: Minimum evaluation set size.}
At least 20 examples per category are required, for a minimum total
of 260 examples across 13 categories.
This provides 95\% Wilson confidence intervals of approximately
$\pm$22\% per class, sufficient for directional conclusions
and significantly better than the $\pm$60\% intervals produced
by 4-example per-class sets.

\textbf{R2: Fuzzy field extraction.}
The primary accuracy metric must use a parser that handles field-name
format variation.
Strict-regex results may be reported as a secondary metric to
quantify parsing-induced suppression, but must not be the primary
figure cited in papers or comparisons.

\textbf{R3: Failure inspection for low-accuracy categories.}
For any category with accuracy below 50\%, the evaluator must report
whether failures are (a) wrong predictions, (b) empty predictions
from generation failure, or (c) extraction failures from the parser.
This prevents pipeline failures from being silently reported as
model failures.

\textbf{R4: Evaluation metadata documentation.}
Published evaluations must report: inference parameters
(\texttt{max\_new\_tokens}, \texttt{temperature}, \texttt{do\_sample});
parser type (strict or fuzzy); normalization rules; and any
post-processing applied to raw output.
Without this information, results cannot be reproduced or compared.

\subsection{Scoring Metric}

The primary SOC-Bench metric is macro-averaged broad-category accuracy:
the unweighted mean of per-class accuracy across all 13 categories.
Macro-averaging weights each category equally regardless of evaluation
set frequency, preventing high-volume categories from dominating
the headline figure.

Secondary metrics include severity accuracy, MITRE technique extraction
rate (for parsers that extract MITRE fields), and per-class
Wilson confidence intervals.

\subsection{Public Artifacts}

SOC-Bench v0 scoring scripts are publicly available at
\url{https://github.com/chaitanyagarware/soc-bench}.
The repository includes: the fuzzy field extractor,
the 13-category normalization rules, the scoring script
(callable as \texttt{python evaluate.py --predictions preds.json
--ground\_truth gt.json}),
and the 50-example evaluation set used in this study.
Researchers can produce SOC-Bench-comparable results by running
their model on the evaluation set and passing outputs to the scoring script.

\subsection{What SOC-Bench Prevents}

SOC-Bench v0 is designed to prevent the following failure modes:

\begin{itemize}
\item \textbf{Parser-specific false failures.}
The fuzzy extraction requirement eliminates the 76pp accuracy gap
documented in this paper from appearing as a model failure.

\item \textbf{Incomparable accuracy figures.}
Shared taxonomy, normalization rules, and scoring logic produce
figures that are comparable across research groups.

\item \textbf{Hidden extraction errors.}
R3 (failure inspection) requires per-category inspection that
catches both extraction failures and generation failures
before they contaminate aggregate accuracy.

\item \textbf{Underpowered per-class claims.}
R1 (minimum evaluation size) ensures that per-class conclusions
are backed by sufficient statistical power to be meaningful.

\item \textbf{Irreproducible evaluations.}
R4 (metadata documentation) ensures that other groups can
reproduce published results without reverse-engineering
undocumented pipeline details.
\end{itemize}

\section{Discussion}
\label{sec:discussion}

\subsection{Evaluation Correctness Is Not an Implementation Detail}

The central lesson of this work is simple but has broad consequences:
in SOC LLM evaluation, the parser is part of the measurement instrument.
If it fails silently, accuracy becomes a property of the evaluator
rather than the model.
A 76 percentage-point accuracy gap produced by a field-name format
mismatch is not a rounding error or an edge case.
It is the difference between concluding that a model is useful
and concluding that it is broken.

This finding is not specific to regex parsers or to TinyLlama.
Any system that extracts structured labels from free-form LLM output
faces the same risk.
The specific format mismatch (underscore vs.\ space in field names)
will vary by model family and fine-tuning configuration,
but the category of failure such as silent suppression of correct predictions
by a brittle extractor is structural.
It will recur whenever evaluation pipelines are not audited against
the actual output format of the model being evaluated.

\subsection{Published Accuracy Figures Should Be Treated as Lower Bounds}

For any LLM-based SOC classifier evaluated with a regex extractor,
the reported accuracy figure is a lower bound on the model's semantic
accuracy until the extraction methodology is independently validated.
The magnitude of suppression is not predictable from code inspection
alone: it depends on the specific formatting preferences of the base
model, which vary across model families, scales, and instruction-tuning
configurations.
In this study the gap was 76 percentage points.
In other systems it may be smaller or larger.
The only way to know is to run a parser comparison.

We recommend that any paper reporting accuracy for an LLM-based
SOC classifier include the following in its evaluation methodology:
the raw output format of the model, the field names it generates,
the parser logic used for extraction, the normalization rules applied,
and a parser comparison (strict vs.\ fuzzy) on at least a sample
of the evaluation set.

\subsection{Small Models May Be Systematically Undervalued}

The result that TinyLlama-1.1B + LoRA achieves 76\% threat accuracy
under correct evaluation is approaching Claude Sonnet's 88\% while
being trained in 5 minutes on 450 examples suggests that
brittle evaluation pipelines may be systematically undervaluing
small fine-tuned models.
A model that appears to fail completely under strict parsing
may be genuinely useful under correct evaluation.
The research community's perception of where small models sit
relative to large zero-shot models may be distorted by accumulated
evaluation artifacts of this type.

This does not mean small models are as capable as large ones.
The 12pp gap between TinyLlama fuzzy and Claude zero-shot is real.
But the gap that actually matters for practical deployment decisions
is 12pp, not 88pp.
Whether a resource-constrained organization should deploy a 1.1B
fine-tuned model depends on whether 76\% accuracy is operationally
useful, not on whether a broken parser reports 0\%.

\subsection{The Cost of Silent Failures}

What makes parsing-induced suppression particularly damaging is
that it produces no error signal.
The evaluation script runs to completion.
Accuracy metrics are computed.
Results are reported.
Everything appears to have worked.
The only way to detect the failure is to inspect raw model outputs
and compare them against what the parser expected to see, a step
that is not standard practice in the field.

This is a systemic problem, not a researcher error.
Current practice does not require parser auditing,
does not require raw output format documentation,
and does not provide shared evaluation infrastructure
that would surface these failures automatically.
SOC-Bench v0 is designed to make the auditing step standard.

\section{Limitations}
\label{sec:limitations}

\textbf{Evaluation set size.}
The 50-example evaluation set provides sufficient power to detect
the parsing-induced suppression effect and identify the failure
concentration pattern, but per-class estimates based on 1--8
examples carry wide confidence intervals.
The 0\% results for Reconnaissance, Brute Force, and Credential
Stuffing are consistent with systematic failure but do not
establish failure rates precisely.
SOC-Bench v0's minimum of 20 examples per class is the
appropriate correction.

\textbf{Fuzzy parser false positives.}
A fuzzy parser that is too permissive can map an incorrect prediction
to the correct canonical category, producing spurious accuracy.
Our implementation mitigates this by requiring the correct category
keyword to be present in the predicted text, but we have not
systematically measured the false-positive rate of this design.
Future work should validate fuzzy parser precision alongside recall.

\textbf{Claude Sonnet baseline reproducibility.}
Claude Sonnet is a proprietary model whose architecture and exact
parameter count are not public.
Results reported here for this baseline may change as the model
is updated, and other research groups cannot reproduce the
exact model state used in this evaluation.
We report these results for contextual reference, not as a
reproducible benchmark.

\textbf{Generation artifacts not causally validated.}
We observed prompt-repetition artifacts in a subset of raw outputs
for the three 0\%-accuracy categories.
We did not conduct controlled experiments varying inference parameters
to establish whether these artifacts causally explain the failures.
We do not claim that fixing generation parameters would produce
100\% accuracy.
This is a hypothesis that requires experimental validation on an
expanded evaluation set.

\textbf{Single model family.}
All fine-tuning experiments use TinyLlama-1.1B.
Whether the same parsing-induced suppression magnitude would appear
with other model families (Mistral, Phi, Gemma) at similar scales
is an open question.
The structural argument is that instruction-tuned models generate
different field name formats than fine-tuning templates applies
broadly, but the specific suppression magnitude will vary.

\section{Conclusion}
\label{sec:conclusion}

We set out to understand why a fine-tuned TinyLlama-1.1B model
appeared to achieve 0\% threat accuracy on a 50-example held-out set.
The answer was not model failure. It was measurement failure.

A strict regex extractor searching for \texttt{THREAT\_TYPE:}
silently discarded every prediction where the model had generated
\texttt{Threat Type:} which is a formatting variant that differs only in
underscore vs.\ space.
Replacing the extractor with a format-tolerant fuzzy parser
recovered 76\% threat accuracy on the same model outputs.
Severity accuracy remained at 58\% under both parsers,
providing a built-in control that confirms the failure was
parser-specific rather than a consequence of model degradation.
Claude Sonnet, evaluated zero-shot on the same set under the
same fuzzy protocol, achieved 88\% threat accuracy, a 12pp gap
from the fine-tuned small model, not the 88pp gap that a strict
parser would have suggested.

Residual errors concentrate in three behaviorally adjacent categories:
Reconnaissance, Brute Force, and Credential Stuffing.
Each contributes 4 misclassifications.
All other categories achieve 100\% accuracy.
This concentration pattern reflects systematic class-boundary
difficulty rather than global model failure, and it motivates
targeted improvement, better boundary examples, larger evaluation
sets, taxonomy alignment rather than architectural change.

We introduce SOC-Bench v0: a benchmark framework with standardized
taxonomy, minimum statistical power requirements, fuzzy evaluation
protocol, and a public scoring script to prevent parser-induced
accuracy distortion in future SOC LLM research.

The broader lesson is methodological.
Evaluation pipelines in LLM-based security research are fragile,
their failures are silent, and their effects on reported accuracy
can be large.
The field needs shared evaluation infrastructure that makes
these failures visible before they reach published results.

Evaluation correctness is not a secondary concern;
it is a first-order determinant of perceived model capability
in SOC LLM systems.

\section*{Acknowledgment}

The authors thank the University of Alabama at Birmingham Department of Computer and Information Sciences for research support.
SOC-Bench scoring scripts are publicly available at
\url{https://github.com/chaitanyagarware/soc-bench}.


\end{document}